\documentclass[a4paper]{jpconf}
\usepackage{fontenc,indentfirst,delarray,amsmath,amssymb}
\usepackage{graphicx,graphics,epstopdf,epsf}
\usepackage{pdfsync}
\usepackage{color}

\def\del{\partial}
\def\lp{\left(} 
\def\rp{\right)} 
\def\dm{\lp\begin{array}}	
\def\fm{\end{array}\rp}
\def\m2{M_2 \lp \cc \rp}
\def\m3{M_3 \lp \cc \rp}

\def\ds{\partial\!\!\!\slash}

\def\cc{{\mathbb{C}}}
\def\C{{\mathbb{C}}}

\def\I{{\mathbb{I}}}

\def\H{{\cal H}}	
\def\mm{{\mathcal M}}	
\def\M{{\mathcal M}}

\def\A{{\mathcal A}}

\def\cinf{C^{\infty}\lp\mm\rp}

\def\xo0{\omega^0_x}
\def\yo0{\omega^0_y}

\def\xo0{x_\omega^0}
\def\yo0{y_\omega^0}

\def\fm{\Phi(x^\mu)}

\def\dm{\partial_\mu}

\begin{document}
\title{Twisted spectral geometry for the standard model}

\author{Pierre Martinetti}

\address{Universit\`a di Trieste, via Valerio 12/1, I-34127}

\ead{pmartinetti@units.it}

\begin{abstract}
The Higgs field is a connection $1$-form as the other bosonic fields, 
provided one describes space no
more as a manifold $\M$ but as a slightly non-commutative generalization of
it. This is well encoded within the theory of spectral triples: all the bosonic fields of the standard model - including the Higgs -
are obtained on the same footing, as fluctuations of a generalized Dirac operator by a matrix-value algebra
of functions on $\M$. In the commutative case, fluctuations of the
usual free Dirac operator by the complex-value algebra $\cinf$ 
of smooth functions on $\M$ vanish, and so do not generate any bosonic
field. We show that imposing a twist in the sense of Connes-Moscovici forces to double the
algebra to $\cinf\otimes \C^2$, but does not require to modify the space of
spinors on which the algebra acts. This opens the way to twisted fluctuations of the
free Dirac operator, that yield a perturbation of the spin connection.
 Applied to the standard model, a similar twist yields  in addition the extra scalar field 
needed to stabilize the electroweak vacuum, and to make the computation
of the Higgs mass in noncommutative geometry compatible with its
experimental value.\\
\emph{Proceedings of DICE 2014 ``Spacetime, matter,
  quantum mechanics'', Castiglioncello sept. 2014.} 
\end{abstract}

\section{Introduction}
Noncommutative geometry \cite{Connes:1994kx} provides a description of the standard model
of elementary particles [SM] as a purely gravitational theory \cite{Connes:1996fu}. By this, one
means  that assuming space~(time) is described by a slightly non-commutative generalization
of a manifold, then Einstein-Hilbert action (in Euclidean signature) together with the bosonic
action of the SM, including the Higgs sector, are obtained from one single action
formula, the \emph{spectral action} \cite{Chamseddine:1996kx}. The bosonic Lagrangian is the noncommutative counterpart
of the Einstein-Hilbert action, and the Higgs field comes out on the
same footing as the other bosons as a connection $1$-form,
but associated to the noncommutative part of the geometry
\cite{Dubois-Violette:1989fk, Connes:1996fu}.

More precisely, the Einstein-Hilbert action together with the
various pieces of the SM are obtained as the asymptotic expansion
$\Lambda\to\infty$ of the spectral action, where $\Lambda$ is a cut-off
parameter. The spectral
action thus provides some boundary conditions between the parameters of
the SM at a putative energy of unification, and physical
predictions are obtained by
running these parameters under the renormalization
group flow down to the electroweak breaking
scale \cite{Chamseddine:2007oz}. 
The mass of the Higgs boson is then function of
the inputs of the theory (mainly the Yukawa coupling of the
fermions and the mixing angles for quarks and neutrinos). Assuming
there is no new physics between the electroweak and the unification
scales, it is computed
around $170 \text{ Gev}$, a value ruled out by Tevatron
in 2008. 

Since then the Higgs boson has been discovered 
around $125\text{ GeV}$, which is below the threshold of stability of 
the electroweak vacuum: the later is a metastable state.
It is not clear whether this is a problem or not, since the life-time
of this metastable state is far larger than the age of the
universe. However the estimation of the probability that somewhere in
our past light-cone the Higgs field has tunneled
down to its true vacuum - liberating a quantity of energy that should
have destroyed the
whole universe - depends on the model of
Inflation at the tip of the light-cone. So even
if not problematic, it is at
least intriguing that 
the electroweak vacuum is metastable, but on the edge of stability (see
\cite{near-critic} for a recent update on these issues).

To cure this instability, a long known solution proposed by particle
physicists is to
postulate another heavy scalar field, say $\sigma$, suitably coupled to the Higgs. 
In noncommutative geometry, bosonic fields are obtained by so
called \emph{fluctuations of the metric}, roughly speaking  
 a way to turn the constant parameters of the theory (the
Yukawa couplings) into fields (there is in reality a intricate
re-parametrization and the correspondance Yukawa couplings/bosonic
fields is more subtle). Chamseddine and Connes noticed in \cite{Chamseddine:2012fk} that by
turning into a field one of the constant entry of the generalized
Dirac operator describing the SM, namely the Majorana mass $k_R$ of the neutrino, then one gets
exactly the field $\sigma$ suitably coupled to the Higgs. As a bonus, $\sigma$ modifies
the flow of the renormalisation group and makes the computation of the
Higgs mass 
compatible with its experimental value. The question is then to
understand how to turn the constant parameter $k_R$ into a field
respecting the framework of noncommutative geometry. The
problem is that unlike the
other bosonic fields,  $\sigma$ cannot be obtained by a fluctuation of the
metric because the constant parameter $k_R$ only fluctuates
to a constant field. 
 This impossibility has its origin in of one mathematical requirements of noncommutative
geometry, namely  the condition asking that the generalized Dirac operator is a
first-order differential operator.

Various models have been proposed to justify the turning
of $k_R$ into a field. In \cite{Chamseddine:2013fk,Chamseddine:2013uq}  the first
order condition is relaxed, yielding to a Pati-Salam generalization of
the standard model. The later is retrieved dynamically, as a minimum
of the spectral action. Earlier models pre-2012
had already shown how to lower the Higgs mass thanks to
extra-scalar fields, but they required also new fermions
\cite{Stephan:2009fk,Stephan:2013fk}. Recently, in
\cite{Farnsworth:2014aa} a variation on the notion of symmetry in NCG
yields a model with extra bosonic and scalar fields carrying a $B-L$
charge; in \cite{DAndrea:2015aa} new fields are obtained as a
consequence of a non-standard grading.

 In \cite{Devastato:2013fk} we proposed to generate
the field $\sigma$ in a way satisfying, at least partially, the first
order condition. The key idea is to allow the commutative algebra
$\cinf$ of smooth functions on a compact spin manifold $\M$ to act non-trivially on the space of spinors $L^2(\M, S)$. The drawback is that the commutator 
\begin{equation}
  \label{eq:23}
  [\ds, f]  \quad f\in\cinf
\end{equation}
with the free Dirac operator $\ds$ is no longer bounded
\cite{Devastato:2014fk}, in contradiction with one of the primary requirements
of noncommutative geometry. There exists however a variation of these requirements, introduced in \cite{Connes:1938fk} to deal precisely
with this kind of problem. Given an algebra $\A$ and
a generalized Dirac operator, rather than the boundedness of $[D,a]$ one asks that there
exists an automorphism $\rho$ of $\A$ such that the \emph{twisted
  commutator}
\begin{equation}
  \label{eq:24}
  [D, a]_\rho:= Da -\rho(a)D
\end{equation}
is bounded. Such twists have mathematical motivations that have
nothing to do with physics, but we showed in \cite{buckley} how a very simple
twist of the model proposed in \cite{Devastato:2013fk} -
in fact a chiral transformation - permits to obtain a coherent picture of the SM in which the field $\sigma$ is generated by a fluctuation of
$k_R$ that satisfies a twisted version of the first-order condition.
\smallskip

In this note we give a non-technical account of these results. Rather
than the up-bottom approach developed in \cite{buckley} and summarized
in \cite{Devastato:2015aa}
(i.e.  twists as solution to the
unboundeness of the commutator coming from the non-trivial action of
$\cinf$ on spinors), we propose a bottom-up approach: after some
generalities in section \ref{section:twisted}, we 
show in section \ref{section:need} how requiring a non-trivial twist forces the manifold to be multiplied by a matrix
geometry . We discuss the physical consequences, in particular the
generation of new fields. Finally in section \ref{sectionSM} we state the results of
\cite{buckley} on the standard model of elementary particles.
\newpage

\section{Twisted almost-commutative geometry}
\label{section:twisted}

\subsection{Almost commutative geometry in a nutshell} By ``slightly noncommutative
generalization of a manifold'', one intends  a ``space'' such that the set of functions defined on it is
no longer commutative, but is of the kind
\begin{equation}
  \label{eq:3}
  \cinf \otimes \A_F
\end{equation}
where $\A_F$ is a finite dimensional algebra. This is called an
\emph{almost-commutative geometry} because the center $\cinf$ of the
algebra \eqref{eq:3} is infinite
dimensional (as an algebra). It is represented on 
\begin{equation}
  \label{eq:8}
  L^2(\M,S) \otimes \H_F
\end{equation}
where $\H_F$ is a finite dimensional space (whose basis are the
fermions of the model) carrying a representation of $\A_F$, while $L^2(\M, S)$ is the
space of spinors on which functions
act by multiplication:
$(\pi(f)\psi)(x) = f(x)\psi(x)$ for any $x\in\M$. That is $f$ is
represented as the operator
\begin{equation}
\pi(f):=f \I
\label{eq:6}
\end{equation}
where $\I$ is the identity matrix of
dimension $n$ the numbers of components of a spinor. 

For the standard model,
$\H_F=\H_{sm}$ has dimension the number of fermions ($2$ colored
  quarks  $+ 1$ electron $+ 1$ neutrino $= 8$
multiplied by $2$ (chirality),  $2$ (antiparticles) and $3$
(generations) $ =96$). The space \eqref{eq:8} is then
the space of fermionic fields of the SM. Notice the overcounting of
degrees of freedom: the distinction between chirality and
anti/particles is taken into account both in the finite dimensional
space $\H_F$ and by the number of components of the spinors in
$L^2(\M,S)$. This \emph{fermion doubling} is projected out on the
fermionic action thanks to a Pfaffian \cite{Chamseddine:2007oz}, thus
it is not a real nuisance,
except maybe from an aesthetic point of view.  From our perspective,
it provides in fact a solution for generating the extra scalar field
$\sigma$ as explained in section \ref{sectionSM}. 

The key idea of noncommutative geometry is that all the
geometrical information of the manifold $\M$ is encoded within the Dirac
operator $\ds= -i\gamma^\mu\nabla_\mu$, where $\gamma^\mu$'s are the Dirac
matrices and $\nabla_\mu:=\del_\mu + \omega_\mu$ is the covariant
derivative associated to the spin connection $\omega_\mu$. Connes worked out a purely algebraic characterization of the
Dirac operator, that he then exported to the noncommutative setting
\cite{Connes:1996fu}. Hence the notion of \emph{spectral
  triple}, that is an algebra $\A$ (non necessarily commutative), acting
through a representation $\pi$ on a Hilbert space $\H$, together with an operator $D$ with
compact resolvent (or a generalization of this condition in case the
algebra is not unital) called (generalized) Dirac operator such that
\begin{equation}
  \label{eq:1}
  ||[D, \pi(a)]|| \text{ is bounded for any } a \in\A.
\end{equation}
One also asks that $\cal H$ is a graded Hilbert space, that is there
exists an operator $\Gamma$ such that $\Gamma^2=1$ which, furthermore,
anticommutes with $D$ and commutes with $\A$.
For the standard model one has
\begin{equation}
  \label{eq:25}
  \Gamma = \gamma^5 \otimes \Gamma_{sm}
\end{equation}
where, writing 
\begin{equation}
  \label{eq:26}
  \H_{sm} = \H_R \oplus \H_L \oplus H_R^c \oplus \H_L^c,
\end{equation}
as the sum of four copies of $\C^{24}$ labelled by left/right and anti/particles
indices, one has $\Gamma_{sm}=\text{diag}\,(\I_{24}, -\I_{24},
-\I_{24},\I_{24})$ while $\gamma^5$ is the product of the Dirac matrices. 

Adding other conditions, one shows that given any spectral
triple with unital commutative algebra $\A$, then there exists a compact spin manifold $\M$ such that
$\A=\cinf$ \cite{connesreconstruct}. The conditions on the analytic
properties of the operator $D$ are automatically satisfied in the
finite dimensional case. As well, we shall not take into account here the
orientability condition and Poincar\'e duality, but will focus on the
\emph{real structure} and the already mentioned \emph{first order
  condition}. The real structure $J$ is an antilinear operator ($J(\lambda \psi) = \bar\lambda J \psi$
for $\lambda\in\C,\, \psi\in \cal H$) whose square is $\pm \I$, and
which commutes or anticommutes with the graduation $\Gamma$ and the
operator $D$. The three signs
\begin{equation}
J^2= \epsilon \I,\;  JD
=\epsilon' DJ,\;  J\Gamma = \epsilon'' \Gamma J
\label{eq:28}
\end{equation}
determines the so called $KO$-dimension of the spectral triple.

All these conditions are satisfied by the triple $(\cinf, L^2(\M,S),
\ds)$, with real structure the charge conjugation $\cal J$, grading
the chirality $\gamma^5$. The $KO$-dimension is then the dimension of the
manifold $\M$ (modulo $8$). A classification of finite dimensional
spectral triples satisfying these conditions as been made in
\cite{Chamseddine:2007oz,Chamseddine:2008uq} and yields an (almost)
unique choice $(\A_{sm}, \H_{sm}, D_{sm})$  relevant for the standard
    model. The choice of the algebra $\A_{sm}$ is discussed in \S
    \ref{sectionSM}, the Hilbert space is $\H_{sm}=\C^{96}$ described below \eqref{eq:6}, the constant entries of $96\times 96$ matrix $D_{sm}$ 
are the Yukawa coupling of fermions and the mixing matrices for quarks and
neutrinos. A general formula of
products of spectral triples yields the almost-commutative spectral
triple 
\begin{equation}
  \label{eq:10}
\cinf\otimes\A_{sm},\quad L^2(\M,S) \otimes \H_{sm},\quad
  D:= \ds \otimes \I + \gamma^5\otimes D_{sm}.
\end{equation}
The graduation is given in \eqref{eq:25} and the real structure is
${\cal J}\otimes J_{sm}$ where $J_{sm}$ is the operators that exchange
particles with antiparticles in $\H_{sm}$.

\subsection{Bosonic fields and spectral action}
Bosonic fields are generated by the so-called \emph{fluctuations of
  the metric}. Given an arbitrary spectral triple $(\A, \H, D)$, those
are defined as the substitution of the
Dirac operator with a \emph{covariant operator} 
\begin{equation}
  \label{eq:21}
  D_A:= D + A + JAJ^{-1}
\end{equation}
where $A$ is a selfadjoint element of the set of (generalized)
$1$-forms
\begin{equation}
  \label{eq:22}
  \Omega^1_D(\A) :=\left\{ \pi(a^i) [D, \pi(b_i)], \quad a^i, b_i\in\A\right\}.
\end{equation}
The name is justified because for $\A_F= M_N(\C)$ and $D_F =0$, then
$D_A$ is nothing but the covariant Dirac operator of a $U(n)$ gauge
theory on $\M$. For the almost-commutative geometry \eqref{eq:10} of the standard
model, these fluctuations generate the bosonic fields and the Higgs.

The spectral actions consists in counting the
eigenvalue of $D_A$ smaller than an energy scale~$\Lambda$,
\begin{equation}
  \label{eq:11}
  S= \text{Tr}\, f\left(\frac{D_A^2}{\Lambda^2}\right)
\end{equation}
where $f$ is a smooth approximation of the characteristic function
of the interval $[0,1]$. As explained in introduction, for the
spectral triple \eqref{eq:10} fluctuated as in \eqref{eq:21}, the
asymptotic expansion of $S$ yields Einstein-Hilbert action and the SM
bosonic action, including the Higgs.

\subsection{Twisted spectral triple}

In the definition \eqref{eq:21} of the covariant Dirac operator, it is important that the commutators $[D,\pi(a)]$ are
bounded, otherwise bosons would be described by unbounded operators. 
Whatever the finite dimensional spectral triple $(\A_F, \H_F, D_F)$,
the commutator $[D_F, \pi(a)]$ is automatically
bounded. The same is true for the
commutative part $[\ds, \pi(f)]=(\ds f)\I$. Hence as long as the almost-commutative algebra
\eqref{eq:3} acts on \eqref{eq:8} with the  trivial action
\eqref{eq:6} on spinors, the
commutator $[D, \pi(a)]$ is always bounded.  

Nevertheless, as explained in \cite{Connes:1938fk}, there are situations where
the requirement \eqref{eq:1} is too strong, like the lift $\ds'$ to $\ds$ of a
conformal map. Then $[\ds',\pi(f)]$ is no longer bounded, but there
exists an automorphism $\rho$ of $\cinf$ such that the twisted commutator 
$[\ds', \pi(f)]_\rho$ is bounded. More generally, requiring the
boundedness of the twisted commutator makes sense mathematically and
allows to treat cases (type III spectral triple) where the
usual commutator is never bounded \cite{Connes:1938fk}. 
This  yields the definition of a \emph{twisted spectral triple}  $(\A,
\cal H, D, \rho)$, similar to a spectral triple
except that $[D,\pi(a)]$ bounded is replaced by $[D,\pi(a)]_\rho$
bounded for some automorphism $\rho$.

To the best of our knowledge, the conditions for the
reconstruction theorem \cite{connesreconstruct} (in particular the real structure) have not
been adapted to the twisted case yet. This does not matter because in the
commutative case $(\cinf, L^2(\M,S), \ds)$ with representation \eqref{eq:6} (which
is the one  adressed by the reconstruction theorem), twists are not
relevant. Indeed, whatever the automorphism $\rho$ of $\cinf$,
{\footnote{One has, omitting the symbol $\pi$ of representation:
    $\nabla_\mu f = \del_\mu f +
    \omega_\mu f =  (\del_\mu f) + f\del_\mu + \omega_\mu f = (\del_\mu f) + f\nabla_\mu$.}}
\begin{equation}
  \label{eq:5}
  [\ds, \pi(f)]_\rho  = -i\gamma^\mu (\del_\mu \pi(f))-i [\gamma^\mu\pi(f)-
  \pi(\rho(f))\gamma^\mu]\nabla_\mu
\end{equation}
is bounded if and only if the differential part  vanishes,
that is
\begin{equation}
[\gamma^\mu, \pi(f)]_\rho =0 \quad \forall f\in\cinf.
\label{eq:14}
\end{equation}
By \eqref{eq:6} this is
equivalent to $f=\rho(f)$ for any $f$, that is $\rho$ is the trivial automorphism. 

\section{Need for twist}
\label{section:need}

The precedent section shows that for the usual spectral
triple of a manifold, not only there is no need for a twist because
the usual commutator is bounded,  but there
is no space for it. One may wonder what the minimal modifications are,
so that to authorize a non-trivial
twist. We already know one answer \cite{Connes:1938fk}: the
lift to the Dirac operator of a
conformal map. But sticking to the idea that physics is
contained within the Dirac operator, we want to  keep the usual Dirac operator
$\ds$ and rather play with the other parameters: the algebra $\cinf$ and its
representation \eqref{eq:6}.
  
\subsection{Degenerate representation and almost-commutative algebra}
\label{section:degenerate}
We  first try a degenerate
representation $\pi(f) = fp$ for some projection $p\neq \I$. Condition \eqref{eq:14}
becomes
\begin{equation}
  \label{eq:7}
 \gamma^\mu f p - \rho(f) p \gamma^\mu = 0 \quad\forall f\in\cinf.
\end{equation}
For $f=\rho(f)=1$, this implies $[\gamma^\mu, p] =0\, \forall\mu$.
Only the multiple of the identity commute with all Dirac matrices,
hence $p=\lambda\I$ for some $\lambda\in\C$. $\pi(1)=\pi(1)^2$ fixes
$\lambda = 1$, and one is back to~\eqref{eq:6}.

A possibility to have a non-trivial twist would be to act with the
automorphism $\rho$
on $p$ rather than on $f$ in \eqref{eq:7}. To understand this better, let  us
assume $\M$ has dimension $4$ to fix notations. In the chiral basis, the Euclidean Dirac matrices are
\begin{equation}
  \label{eq:18}
  \gamma^\mu = \left(\begin{array}{cc} 0_2 & \sigma^\mu\\ 
     \tilde\sigma^\mu & 0_2\end{array}\right) 
\end{equation}
where $\sigma^\mu =\left\{\I, -i\sigma_i\right\}$, $\tilde\sigma^\mu
  =\left\{\I, i\sigma_i\right\}$ with $\sigma_{i=1,2,3}$, the Pauli
matrices. Consider the representation of the algebra $\C$ on the
Hilbert space $\C^2$ as $\C\ni\lambda\to \lambda\I_2$. Let $\rho$ be the automorphism of $\C^2$ that
permutes the two terms,
\begin{equation}
  \label{eq:19}
  \rho(\lambda_1, \lambda_2) = (\lambda_2, \lambda_1) \quad\quad \forall (\lambda_1, \lambda_2)\in
  \C^2.
\end{equation}
For $\pi$ the representation of $\C^2$ on $\C^4$ given by
$\pi(\lambda_1, \lambda_2) = \lambda_1\I_2 \oplus \lambda_2\I_2$, one has
\begin{equation}
  \label{eq:20}
  [\gamma^\mu, \pi(\lambda_1, \lambda_2)]_\rho = \left(
\begin{array}{cc}
      0_2 & \left[\sigma^\mu, \lambda_2 \I_2\right]\\ 
      \left[{\tilde\sigma}^\mu, \lambda_1\I_2\right]& 0_2
\end{array}\right) = 0 \quad \forall \lambda_1, \lambda_2\in\C^2.
\end{equation}

This suggests that if one could work with two representations $\pi(f)= f
p$,  $\pi'(f):= fp'$  of $\cinf$ where $p,p'$ are two orthogonal projections in $L^2(\M,S)$, then the algebra isomorphism
$ \tau: \pi(f)\to \pi'(f)$
would define a modified-commutator
\begin{equation}
  \label{eq:30}
  \ds \pi(f) - \tau(\pi(f)) \ds =-i\gamma^\mu (\nabla_\mu pf) -i f(\gamma^\mu p - p'
  \gamma^\mu)\nabla_\mu.
\end{equation}
This is
bounded iff the second term vanishes, for instance when $p=\text{diag}\,(\I_2, 0_2)$, $p'=\text{diag}\,(0_2, \I_2)$. 
The point is that $\tau$ is not an
automorphism of $\cinf$, since the algebra
generated by $\pi(f)$ and $\pi'(f)$ for $f\in\cinf$ is not $\cinf$,
but two copies of it, that is $\cinf\otimes\C^2$. 

In other terms,  to have a
non-trivial twist one needs to ``double'' the manifold by multiplying it
by $\C^2$, or more generally by multiplying it by a matrix
algebra $\A_F$. In this sense,  a ``raison d'\^etre'' of almost-commutative
algebra is to allow
non-trivial twists, which are forbidden in the case $(\cinf, L^2(\M,S), \ds)$.

\subsection{Twisted fluctuation of the free Dirac operator}

Fluctuating the free Dirac operator $\ds$ by
$\cinf$ acting as in \eqref{eq:6} has no interest, because~\cite{Landi:1997zm}
\begin{equation}
 \ds_A = \ds.\label{eq:37}
 \end{equation}
The twist introduced in \S \ref{section:degenerate} allows non-trivial fluctuations of $\ds$. Specifically, we
consider 
$\A=\cinf\otimes\A_F$ acting on $L^2(\M,S)$ in such a way that there 
exists an automorphism
$\rho$ of $\A$ guaranteeing that  (omiting the symbol of representation)
\begin{equation}
  \label{eq:27}
 [\gamma^\mu, a]_\rho:=  \gamma^\mu a - \rho(a)\gamma^\mu  = 0 \quad \forall \mu.
\end{equation}
We then define the \emph{twisted-covariant free Dirac operator}
\begin{equation}
  \label{eq:4}
  \ds_{A_\rho}:= \ds + A_\rho + {\cal  J} A_\rho {\cal J}^{-1}
\end{equation}
where $A_\rho$ is a element of the set of twisted $1$-forms
\begin{equation}
  \label{eq:31}
  \Omega_{\ds,\rho}^1(\A) :=\left\{ a^i[\ds, b_i]_\rho,\; a^i, b_i\in \A\right\}.
\end{equation} 
By \eqref{eq:27}  one has
\begin{equation}
[\ds, a]_\rho = -i\gamma^\mu(\nabla_\mu a) - i[\gamma^\mu, a]_\rho
\nabla_\mu = -i\gamma^\mu(\nabla_\mu a)
\label{eq:44}
\end{equation}
where $(\nabla_\mu a)
:= (\del_\mu
  a) + [\omega_\mu,a].$
So a twisted $1$-form is 
\begin{equation}
A_\rho =-i a^i\gamma^\mu(\nabla_\mu b_i) =-i \gamma^\mu X_\mu
\end{equation}
 where
\begin{equation}
X_\mu:=\rho^{-1}(a^i)(\nabla_\mu b_i).
\end{equation}

The claim is that unlike \eqref{eq:37}, $\ds_{A_\rho}$ is not
necessary equal to $\ds$. To see it,
let us consider the simplest example $\A_F=\C^2$ 
acting as in
\eqref{eq:20} with 
$\rho=\rho^{-1}$ as in \eqref{eq:19}. For any  $a^i=(f^i, g^i)$, $b_i=(f'_i, g'_i)$ in $\A=\cinf\otimes
 \C^2$ one has 
\begin{equation}
  \label{eq:40}
  \rho(a^i)=\left( \begin{array}{cc} g^i\,\I_2 & 0_2 \\ 0_2 &f^i\,
      \I_2\end{array}\right), \quad
  b_i=\left( \begin{array}{cc} f'_i\,\I_2 & 0_2 \\ 0_2
      &g'_i\,\I_2\end{array}\right),\quad  X_\mu= \left( \begin{array}{cc}
      g^i(\del_\mu f'_i)\,\I_2 & 0_2 \\ 0_2
      &f^i(\del_\mu g'_i)\,\I_2\end{array}\right)
\end{equation}
where we use that
\begin{equation}
\omega_\mu=-\frac 14 \Gamma_{\mu\rho}^\nu
\gamma^\rho\gamma_\nu = -\frac 14  \Gamma_{\mu\rho}^\nu\left(\begin{array}{cc}
\sigma^\rho\tilde\sigma_\nu &0 \\ 0& \tilde\sigma^\rho\sigma_\nu\end{array}\right)
\label{eq:46}
\end{equation}
 ($\Gamma$ the Christoffel symbol
in the orthonormal basis) commutes with $b_i$. The $KO$-dimension
coincides with the metric dimension of the manifold, that is $4$, meaning that
$\cal J$ commutes with $\ds$ so that 
$[{\cal J}, i\gamma^\mu]=0$. Hence 
\begin{equation}
{\cal J} A_\rho {\cal J}^{-1}  = -i
\gamma^\mu {\cal J} X_\mu{\cal J}^{-1}.
\label{eq:45}
\end{equation}
Furthermore,
\begin{equation}
 {\cal J}
X_\mu{\cal J}^{-1}=  {\cal J} \rho(a^i){\cal J}^{-1}
{\cal J}(\del_\mu b_i){\cal J}^{-1} = \rho(a^i)^* \,(\del_\mu b_i^*) = X_\mu^*
\label{eq:39}
\end{equation}
because ${\cal J} \rho(a) {\cal
  J}^{-1}= \rho(a)^*$ commutes with ${\cal J}(\del_\mu b) {\cal
  J}^{-1} = \del_\mu b^*$. Therefore
\begin{equation}
A_\rho + {\cal J} A_\rho {\cal J}^{-1} = -i\gamma^\mu(X_\mu + X_\mu^*).
\label{eq:36}
\end{equation}
Since $X_\mu$, $X_\mu^*$ twisted-commute with
$\gamma^\mu$,  the adjoint of \eqref{eq:36} is $i\gamma^\mu\rho(X_\mu^*
+X_\mu)$ and $\ds_{A_\rho}$ is selfadjoint as soon as
\begin{equation}
X_\mu + X_\mu^* = -\rho(X_\mu^* + X_\mu).\label{eq:41}
\end{equation}
Writing $f_\mu$, $g_\mu$  the real part of  $g^i\del_\mu f'_i$,
$f^i\del_\mu g'_i$,  so that
  \begin{equation}
    \label{eq:42}
     X_\mu+ X_\mu^*= \left( \begin{array}{cc}
      2f_\mu\,\I_2 & 0_2 \\ 0_2
      &2g_\mu\,\I_2\end{array}\right),
  \end{equation}
condition \eqref{eq:41} is equivalent to $g_\mu =- f_\mu$. Hence 
\begin{equation}
    \label{eq:35}
    \ds_{A_\rho} =\ds - 2i \gamma^\mu\left(\begin{array}{cc} f_\mu \I_2
        &0_2 \\ 0_2&-f_\mu \I_2\end{array}\right)
  \end{equation}
is not necessarily equals to $\ds$. 

This simple example shows that a non-trivial twist has
interesting physical consequences: while fluctuations of the of the free Dirac
operator  by $\cinf$ are trivial, twisted fluctuations by
$\cinf\otimes \C^2$ generate a vector field $X_\mu$. The difference
between the twisted and un-twisted cases is clear from \eqref{eq:36} : if $X_\mu$ were commuting with the Dirac
matrices, then $A_\rho + {\cal J} A_\rho{\cal J}^{-1}$
would be selfadjoint iff $X_\mu+ X_\mu^* = - (X_\mu +
X_\mu^*)$, that is $X_\mu = X_\mu^*$, hence  $A_\rho + {\cal J}
A_\rho{\cal J}^{-1}$ would be zero and $\ds_\rho$ would equal $\ds$. 

The physical
 interpretation of the field $X_\mu$ is delicate:  by making functions acting
 non-trivially on spinors, one breaks the invariance of the representation of $\cinf$ under the spin
 group.
 In this sense, these models are ``pre-geometric'':
 the spin structure is not
 explicit in the representation, but is somehow ``hidden'' in the Dirac operator. 
 Comparing \eqref{eq:35} with \eqref{eq:46}, the field $X_\mu$ appears
 as a perturbation of
the spin connection. It would be interesting to understand whether the
spin connection itself could be generated by a twisted-fluctuation
of the flat Dirac operator.
This has to be put in contrast with almost
commutative geometries, where $\cinf$ acts on spinors as in
\eqref{eq:6} while $\A_F$ acts on a finite dimensional space $\cal
H_F$. Then a (non-twisted) fluctuations of $\ds\otimes {\I}_F$ by
$\cinf\otimes\A_F$ yields a $U(\A_F)$ connection, but the spin
connection is untouched. As explained in the next section, the twisted
spectral triple of the Standard Model combines these two aspects: a
non-trivial action of $\cinf$ on spinors together with an action of $\A_F$ on a
finite dimensional Hilbert space.

One may also be puzzled by our lack of care in viewing  $L^2(\M,S)$ as $L^2(\M)\otimes \C^4$. We argue in
\cite{Devastato:2013fk} that this makes sense in a local trivialization. Eventual non-local
effect  should be studied.

\section{Twist for the standard model}
\label{sectionSM}

Let us conclude by the applications to the standard model. We work
with one generation only, so that the
finite dimensional algebra $\A_{sm}$ (discussed below) acts on the finite dimensional
space $\H_{sm}=\C^{96/3=32}$.
 
\subsection{Grand symmetry}

Independent considerations on the signature of the metric \cite{Barrett:2007vf},
the mass of neutrinos and the fermion doubling \cite{connesneutrino}
indicate that the
$KO$-dimension of the finite dimensional part of the spectral triple
of the SM should be $6$. Under natural hypothesis (irreducible actions
of the algebra and of the real structure, existence of a separating
vector) and an explicit ad-hoc ``symplectic hypothesis'', it has been
shown in \cite{Chamseddine:2008uq} that in order to accomodate a real
structure $J$ and a non-trivial grading $\Gamma$, the finite
dimensional algebra of the almost-commutative geometry of the SM
has to be
\begin{equation}
M_a(\mathbb H) \oplus M_{2a}(\C)
\label{eq:15}
\end{equation}
for $a$ an integer greater than $1$, acting on a space with dimension $d=2(2a)^2$. 
For $a=2$ the dimension $d=32$ is precisely the number of particles
per generation of the SM. By further imposing the grading condition
($[\Gamma,a] = 0$) and the first order condition
($[[D,a], Jb^*J^{-1}]=0$), one arrives to the algebra of the standard model
\begin{equation}
\A_{sm}:=\C\oplus{\mathbb H}\oplus M_3(\C). 
\label{eq:16}
\end{equation}

 In \cite{Devastato:2013fk} we noticed that
for $a=4$, the dimension $d=128$ was precisely $4$ times the number of
particles per generation. Viewing $4$ as the dimension of spinors on a
four-dimensional space (time) $\M$, one identifies  
locally  $L^2(\M,S)\otimes \H_{sm}$  as $L^2(\M)\otimes
(\H_{sm}\otimes\C^4)$, which provides precisely the space needed to represent
the \emph{grand algebra} 
\begin{equation}
  \label{eq:13}
  \A_G := M_4(\mathbb H)\oplus M_8(\C).
\end{equation}
Any element $Q$ of  $M_4(\mathbb H)$ and $M$ of  $M_8(\C)$ are viewed as $2\times 2$
matrices with entry in $M_2(\mathbb H)$ and $M_4(\C)$. These entries
act on $\H_{sm}$ as does \eqref{eq:15} for $a=2$, so that at the end of
the game one retrieves the action of $\A_{sm}$. The novelty is that
the $2\times 2$ matrices $Q, M$ have a non trivial action on the remaining
$\C^4$. In this way, one obtains a
representation of $\cinf\otimes\A_G$ with a non-trivial action on
spinors asin \S \ref{section:need}.

The grading condition breaks $\A_G$ to
\begin{equation}
  \label{eq:17}
  {\mathbb H}_L^l\oplus   {\mathbb H}_R^l \oplus {\mathbb
    H}_L^r\oplus   {\mathbb H}_L^r\oplus M_4(\C)
\end{equation}
where $L,R$ are the left-right indices in $\H_F$ and $l,r$ the
left-right indices of spinors.
Because of the non-trivial action on spinors, the commutator
$[\ds\otimes \I, a]$ is never bounded. But there exists a
twist $\rho$ such that $[\ds\otimes \I, a]_\rho$ is bounded, this is simply
the exchange of the spinorial left-right indices:
\begin{equation}
  \label{eq:32}
    {\mathbb H}_L^l\Longleftrightarrow     {\mathbb H}_L^r,\quad
    {\mathbb H}_R^l\Longleftrightarrow {\mathbb
    H}_R^r.
\end{equation}
By further considering the the twisted
version of the first order condition
\begin{equation}
  \label{eq:34}
 [[D, a]_\rho, J b^*J^{-1}]_\rho = 0 
\end{equation} for the Dirac operator of the SM
\begin{equation}
D= \ds\otimes
\I + \gamma_5\otimes D_{sm},
\label{eq:12}
\end{equation}
one works out a sub-algebra of \eqref{eq:17} acting on
$L^2(\M,S)\otimes (\C^4\otimes \H_{sm})$, namely
\begin{equation}
  \label{eq:33}
  {\cal B} :=   {\mathbb H}_L^l\oplus   {\mathbb C}_R^l \oplus {\mathbb
    H}_L^r\oplus   {\mathbb C}_R^r\oplus M_3(\C)
\end{equation}
which, together
with $D$, defines a twisted spectral triple \cite{buckley}. 

 \subsection{Extra scalar field and additional vector fields}
Twisted fluctuations of $\ds\otimes \I$ by $\cal B$  yields a vector
field $X_\mu$ as in \eqref{eq:40}. Twisted fluctuations of
$\gamma^5\otimes D_R$, where $D_R$ the part of $D_{sm}$ containing the Majorana mass $k_R$ of
the neutrino, generates a scalar field $\boldsymbol\sigma$ which coincides with
the field $\sigma$ studied in \cite{Chamseddine:2008uq} up to a global
$\gamma^5$ factor. The spectral action yields a potential for these two fields, which is minimum precisely
when $D$ is fluctuated by the sub-algebra of $\cinf\otimes\cal B$ invariant under
the twist, that is by $\cinf\otimes\A_{sm}$~\cite{buckley}. 

\section{Conclusion}
 The twist \eqref{eq:32} allows to build a twisted spectral triple for
 the standard model with a non-trivial action on spinors.  The
 extra scalar field $\boldsymbol\sigma$ is generated by a
 twisted-fluctuation of the Majorana part of the Dirac operator, while
 twisted-fluctuations of the free Dirac operator generate an
 additional-vector field. All these fluctuations satisfy a twisted
 version of the first order condition. Furthermore, as in
 \cite{Chamseddine:2013uq} the breaking to the standard model (in our
 cae: the un-twisting) is obtained dynamically by minimizing the
 spectral action.
\bigskip

\noindent{\bf Acknowldegments}: Thanks to G. Landi for discussion. Supported by the italian ``Prin 2010-11 Operator Algebras, Noncommutative.
    Geometry and Applications''.

\section*{References}
\bibliographystyle{amsplain}
\bibliography{/Users/pierre/physique/articles/Bibdesk/biblio}

\providecommand{\bysame}{\leavevmode\hbox to3em{\hrulefill}\thinspace}
\providecommand{\MR}{\relax\ifhmode\unskip\space\fi MR }
\providecommand{\MRhref}[2]{%
  \href{http://www.ams.org/mathscinet-getitem?mr=#1}{#2}
}
\providecommand{\href}[2]{#2}
\begin{thebibliography}{10}

\bibitem{Barrett:2007vf}
John~W. Barrett, \emph{A {L}orentzian version of the non-commutative geometry
  of the standard model of particle physics}, J. Math. Phys. \textbf{48}
  (2007), 012303.

\bibitem{near-critic}
D.~Buttazzo, G.~Degrassi, P.~P. Giardino, G.~F. Giudice, F.~Sala, and
  A.~Salvio, \emph{Investigating the near-criticality of the {H}iggs boson},
  arXiv:1307.3536 [hep-ph].

\bibitem{Chamseddine:1996kx}
A.~H. Chamseddine and A.~Connes, \emph{The spectral action principle}, Commun.
  Math. Phys. \textbf{186} (1996), 737--750.

\bibitem{Chamseddine:2008uq}
\bysame, \emph{Why the standard model ?}, J. Geom. Phys \textbf{58} (2008),
  38--47.

\bibitem{Chamseddine:2012fk}
\bysame, \emph{Resilience of the spectral standard model}, JHEP \textbf{09}
  (2012), 104.

\bibitem{Chamseddine:2007oz}
A.~H. Chamseddine, A.~Connes, and M.~Marcolli, \emph{Gravity and the standard
  model with neutrino mixing}, Adv. Theor. Math. Phys. \textbf{11} (2007),
  991--1089.

\bibitem{Chamseddine:2013uq}
A.~H. Chamseddine, A.~Connes, and Walter van Suijlekom, \emph{Beyond the
  spectral standard model: emergence of {P}ati-{S}alam unification}, JHEP
  \textbf{11} (2013), 132.

\bibitem{Chamseddine:2013fk}
\bysame, \emph{{I}nner fluctuations in noncommutative geometry without first
  order condition}, J. Geom. Phy. \textbf{73} (2013), 222--234.

\bibitem{Connes:1996fu}
A.~Connes, \emph{Gravity coupled with matter and the foundations of
  noncommutative geometry}, Commun. Math. Phys. \textbf{182} (1996), 155--176.

\bibitem{connesneutrino}
\bysame, \emph{Noncommutative geometry and the standard model with neutrino
  mixing}, JHEP \textbf{081} (2006).

\bibitem{Connes:1938fk}
A.~Connes and H.~Moscovici, \emph{Type {III} and spectral triples}, Traces in
  number theory, geometry and quantum fields, Aspects Math. Friedt. Vieweg,
  Wiesbaden \textbf{E38} (2008), 57--71.

\bibitem{Connes:1994kx}
Alain Connes, \emph{Noncommutative geometry}, Academic Press, 1994.

\bibitem{connesreconstruct}
\bysame, \emph{On the spectral characterization of manifolds}, J. Noncom. Geom.
  \textbf{7} (2013), no.~1, 1--82.

\bibitem{DAndrea:2015aa}
F.~D'Andrea and L.~Dabrowski, \emph{The standard model in noncommutative
  geometry and {M}orita equivalence}, arXiv:1501.00156 [math-phys] (2015).

\bibitem{Devastato:2015aa}
A.~Devastato, \emph{Noncommutative geometry, grand symmetry and twisted
  spectral triple}, arXiv 1503.03861 (2015).

\bibitem{Devastato:2013fk}
A.~Devastato, F.~Lizzi, and P.~Martinetti, \emph{{G}rand {S}ymmetry, {S}pectral
  {A}ction and the {H}iggs mass}, JHEP \textbf{01} (2014), 042.

\bibitem{Devastato:2014fk}
\bysame, \emph{Higgs mass in noncommutative geometry}, Fortschritte der Physik
  \textbf{62} (2014), no.~9-10, 863--868.

\bibitem{buckley}
A.~Devastato and P.~Martinetti, \emph{Twisted spectral triple for the standard
  and spontaneous breaking of the grand symmetry}, arXiv 1411.1320 [hep-th].

\bibitem{Dubois-Violette:1989fk}
M.~Dubois-Violette, J.~Madore, and R.~Kerner, \emph{Classical bosons in a
  noncommutative geometry}, Class. Quantum Grav. \textbf{6} (1989), 1709.

\bibitem{Farnsworth:2014aa}
S.~Farnsworth and L.~Boyle, \emph{Rethinking {C}onnes' approach to the standard
  model of particle physics via non-commutative geometry}, NJP (2014).

\bibitem{Landi:1997zm}
Giovanni Landi, \emph{An introduction to noncommutative spaces and their
  geometry}, Lecture Notes in Physics m51 (1997).

\bibitem{Stephan:2009fk}
C.~A. Stephan, \emph{New scalar fields in noncommutative geometry}, Phys. Rev.
  D \textbf{79} (2009), 065013.

\bibitem{Stephan:2013fk}
\bysame, \emph{Noncommutative geometry in the {LHC}-era},  (2013).

\end{thebibliography}

\end{document}